\documentclass[twocolumn]{aastex6}
\pdfoutput=1 
\usepackage[T1]{fontenc}
\usepackage{amsmath,amstext}
\usepackage{apjfonts} 
\usepackage[figure,figure*]{hypcap}
\usepackage{microtype}
\usepackage{caption}
\newcommand{\repeater}{FRB\,121102}
\newcommand{\microJy}{\ensuremath{\mu \rm{Jy}}}

\shorttitle{Associating Fast Radio Bursts with Extragalactic Radio Sources}
\shortauthors{Eftekhari et al.}

\begin{document}
\title{Associating Fast Radio Bursts with Extragalactic Radio Sources:  General Methodology and a Search for a Counterpart to FRB 170107}
\author{T.~Eftekhari, E.~Berger, P.~K.~G.~Williams, P.~K.~Blanchard}
\affil{Harvard-Smithsonian Center for Astrophysics, 60 Garden Street, Cambridge, Massachusetts 02138, USA}

\begin{abstract}

The discovery of a repeating fast radio burst (FRB) has led to the first precise localization, an association with a dwarf galaxy, and the identification of a coincident persistent radio source. However, further localizations are required to determine the nature of FRBs, the sources powering them, and the possibility of multiple populations. Here we investigate the use of associated persistent radio sources to establish FRB counterparts, taking into account the localization area and the persistent source flux density. Due to the lower areal number density of radio sources compared to faint optical sources, robust associations can be achieved for less precise localizations as compared to direct optical host galaxy associations. For generally larger localizations which preclude robust associations, the number of candidate hosts can be reduced based on the ratio of radio-to-optical brightness. We find that confident associations with sources having a flux density of $\sim 0.01-$1 mJy, comparable to the luminosity of the persistent source associated with FRB\,121102 over the redshift range $z \approx 0.1 - 1$, require FRB localizations of $\lesssim 20''$. We demonstrate that even in the absence of a robust association, constraints can be placed on the luminosity of an associated radio source as a function of localization and DM. We find that for DM $\approx 1000 \rm \ pc \ cm^{-3}$, an upper limit comparable to the luminosity of the FRB\,121102 persistent source can be placed if the localization is $\lesssim 10''$. We apply our analysis to the case of the ASKAP FRB\,170107, using optical and radio observations of the localization region. We identify two candidate hosts based on a ratio of radio-to-optical brightness of $\gtrsim 100$. We find that if one of these is indeed associated with FRB\,170107, the resulting radio luminosity ($1 \times 10^{29} - 4 \times 10^{30} \ \rm erg \ s^{-1} \ Hz^{-1}$, as constrained from the DM value) is comparable to the luminosity of the FRB\,121102 persistent source ($2 \times 10^{29} \ \rm erg \ s^{-1} \ Hz^{-1}$).

\end{abstract}
\keywords{radio continuum: transients -- methods: statistical}

\section{Introduction}\label{sec:intro}

Fast radio bursts (FRBs) are extremely bright, millisecond-duration pulses of coherent radio emission, with large dispersion measures (DMs) that exceed typical Galactic values, hence pointing to an extragalactic origin. Since the discovery of the first FRB in archival data \citep{Lorimer2007}, roughly 30 additional FRBs have been detected, both in archival and real-time searches (\citealt{Keane2012}; \citealt{Thornton2013}; \citealt{Spitler2014}; \citealt{Champion2016}; \citealt{Petroff2016}; \citealt{Caleb2017}). Despite a growing number of FRB detections, the lack of precise localizations (typically $\sim 10^3 \ \rm arcmin^2$) has led to a wide range of suggested progenitor systems: giant pulses and magnetar flares (\citealt{Cordes2016}; \citealt{Lyutikov2016}; \citealt{Popov2013}' \citealt{Lyubarsky2014}), mergers of compact objects (\citealt{Zhang2016}; \citealt{Wang2016}), the collapse of supramassive neutron stars to black holes (\citealt{Falcke2014}), and emission from rapidly spinning magnetars (\citealt{Metzger2017}; \citealt{Kumar2017}). 

The discovery of the repeating FRB\,121102 (\citealt{Spitler2014}; \citealt{Spitler2016}) led to the first precise localization \citep{Chatterjee2017} and the identification of a low metallicity dwarf star forming host galaxy at $z=0.193$ \citep{Tendulkar2017}, with properties remarkably similar to the host galaxies of both long-duration gamma-ray bursts (LGRBs) and hydrogen-poor superluminous supernovae (SLSN) \citep{Metzger2017}. Radio observations with the European VLBI Network (EVN) revealed the presence of a compact, persistent radio source with a projected linear size of $\lesssim 0.7$ pc, co-located with the bursts to $\lesssim 12$ mas \citep{Marcote2017}. The large angular offset between the radio source and the optical center of the galaxy (170$-$300 mas; \citealt{Tendulkar2017}) generally argue against an active galactic nuclei (AGN) origin, although this premise is less clear in the case of dwarf galaxies. Similarly, the resulting ratio of radio-to-optical brightness for the persistent source and host galaxy ($L_{\rm rad}/L_{\rm opt}\approx 104$) is much larger than expected for a star formation origin. This has led to the suggestion that the radio source may be a nebula associated with FRB\,121102 \citep{Metzger2017}. More recently, the discovery of a large rotation measure with $\sim 10\%$ variations on half-year timescales suggests the presence of a dynamic and highly magnetized environment surrounding the source of the FRB \citep{Michilli2018}.

The properties of the host galaxy, the repeating nature of the bursts, and the persistent radio source are consistent with a model in which FRBs are powered by young, millisecond magnetars (\citealt{Metzger2017}; see also \citealt{Piro2016}; \citealt{Murase2016}), the same engines that have been argued to power SLSN and perhaps LGRBs (\citealt{Nicholl2017a}; \citealt{Nicholl2017b}). In this model, individual bursts are emitted via the dissipation of rotational or magnetic energy, while the quiescent radio emission is due to the magnetar wind nebula or the shock interaction between the supernova ejecta and the surrounding circumstellar medium (\citealt{Metzger2017}; \citealt{Beloborodov2017}). 

In this broad scenario, we expect that FRBs should be located preferentially in dwarf galaxies \citep{Nicholl2017c} and be coincident with quiescent radio sources similar to the one observed in FRB\,121102 \citep{Metzger2017}. \citet{Nicholl2017c} demonstrate that $\lesssim 10$ FRB localizations will be sufficient to test various formation channels. Given the high event rate for FRBs ($\sim 600 \ \rm sky^{-1} \ day^{-1}$ above 1 Jy; \citealt{Lawrence2017}) and the improved sensitivities of upcoming telescopes and surveys, the number of detected bursts is expected to increase drastically over the next few years as new radio facilities come on-line. However, due to a wide range of localization capabilities, precise localizations will continue to be a challenge. 

In \citeauthor{Eftekhari2017} (\citeyear{Eftekhari2017}; hereafter, Paper I), we explored how robustly FRBs with different localization regions can be associated with host galaxies based on the optical brightness of the galaxy. Due to the large areal number density of faint optical sources, we showed that sub-arcsecond localizations are required for confident associations with dwarf galaxies at $z \gtrsim 0.1$, whereas localizations of up to $3''$ may suffice if the hosts are instead generally $L^*$ galaxies, where $L^*$ denotes the characteristic luminosity of a bright galaxy.

Here we explore the likelihood of constraining similar associations with radio sources, motivated by the persistent radio source associated with FRB\,121102 and assuming that all FRBs are coincident with such sources, potentially with a resulting high radio-to-optical flux ratio. We also explore what limits can be placed on the presence and luminosity of a persistent radio source (as a function of localization area) even in the absence of a robust association. We assess our results in the context of existing and upcoming radio facilities. As an example of our method, we present radio (VLA) and optical (Magellan) observations of the localization region of the ASKAP FRB\,170107 \citep{Bannister2017}. We search for radio counterparts in the region using the radio-to-optical flux ratio to identify potential candidates.

The paper is structured as follows. We begin by using radio source number counts to determine the probability of chance coincidence for an FRB and a persistent radio source as a function of localization region and persistent source flux density (\S\ref{sec:nc}). Next, in \S\ref{sec:ullim} we use these results to constrain the upper limit on the radio luminosity of a source in the absence of a robust association. In \S\ref{sec:cands}, we discuss various techniques which may further aid in the identification of associated counterparts. We present an example of our approach in \S\ref{sec:example}, using radio and optical observations of the localization region of FRB\,170107. Finally, we discuss our results in the context of existing and planned FRB search facilities in \S\ref{sec:conc}.

\begin{figure*}
\includegraphics[width=\textwidth]{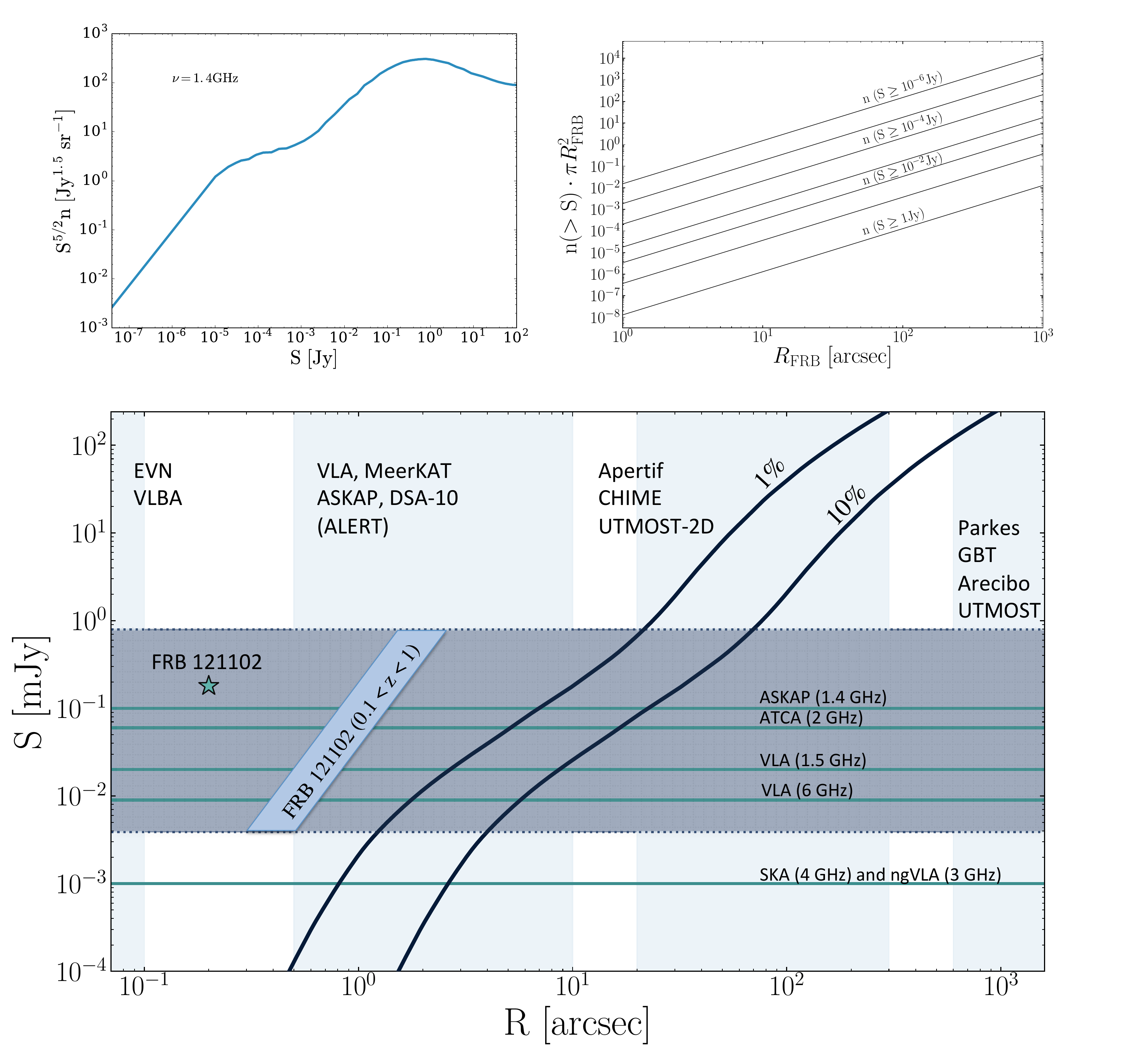}
\caption{\textit{Top-Left:} Euclidean-normalized differential radio source counts (including contributions from AGN and star-forming galaxies) at $1.4$ GHz \citep{Condon2012}. \textit{Top-Right:} The number of radio sources brighter than a limiting flux density $S$ [$n(> S)\pi R^2_{\rm FRB})$] as a function of FRB localization radius ($R_{\rm FRB}$). Individual lines correspond to order of magnitude increments, spanning from 1 $\mu$Jy to 1 Jy. \textit{Bottom:} Probability contours for $P_{\rm cc}=0.01$ and 0.1 as a function of flux density ($S$) and radius ($R$). We also plot the VLA localization and flux density for the quiescent radio source associated with \repeater. Vertical bars indicate the localization regimes of various radio telescopes that are designed to or capable of detecting FRBs (Table 1). Localizations below $20''$ are required for robust associations with radio sources below 1 mJy. Also shown as horizontal lines are the 3$\sigma$ sensitivities for a number of follow-up facilities (assuming 1 hour observations for each).}
\label{fig:nc}
\end{figure*}

\section{Assessing Probability of Chance Coincidence with Radio Source Counts}\label{sec:nc}
We assess the likelihood that an FRB is associated with a persistent radio source by calculating the probability of chance coincidence for a source of a given brightness within a localization region. We use the 1.4 GHz Euclidean-normalized differential source counts presented in \citet{Condon2012} to calculate the number density of radio sources above some limiting flux density, $n(\geq S)$; see top left panel of Figure~\ref{fig:nc}. These source counts are measured over the range $10 \ \mu$Jy $-$ 100 Jy and extrapolated to lower flux densities imposing an evolutionary model for the spectral luminosity function of extragalactic radio sources (\citealt{Condon1984a}; \citealt{Condon1984b}). The source counts are comprised of both AGN and star forming galaxies, with the latter dominating the counts below $\sim 0.1$ mJy \citep{Padovani2015}, although radio-quiet AGN have been shown to comprise a sizable fraction of the sub-mJy population (\citealt{Fomalont2006};  \citealt{Padovani2007}; \citealt{Smolcic2008}; \citealt{Padovani2015}). We assume a Poisson distribution of radio sources across the sky and calculate the chance coincidence probability as: 
\begin{equation} 
P_{\rm cc} = 1 - e^{-\pi R^2 n(\geq S)}.
 \end{equation}
The localization region, $R$, is parameterized by $R=2R_{\rm FRB}$, where $R_{\rm FRB}$ is the $1\sigma$ localization radius of the FRB\footnote{We note that we assume uniform sensitivity across an area with radius $R_{\rm FRB}$. In practice, this implies that the localization region has been uniformly imaged via mosaicking, for instance. In Appendix A, we repeat the analysis presented here, taking into account the primary beam response of the telescope. We find that the assumption of uniform sensitivity is justified, with a more accurate beam model resulting in negligible differences.}.

Using the integrated source counts, we plot in the top right panel of Figure~\ref{fig:nc} the number of radio sources above some limiting flux density [i.e., $\pi R_{\rm FRB}^2n(\geq S$)] as a function of $R_{\rm FRB}$. We plot the results over the range $S = 1 \ \mu$Jy to 1 Jy, in order of magnitude increments. We find an expectation value of about one source at $S \approx 100$ mJy within the typical localization regions of single-dish telescopes ($R_{\rm FRB} \approx 10'$). Conversely, at the faint end ($\approx 0.1$ mJy), we expect one source within $R_{\rm FRB} \approx 1'$. 

We plot the resulting chance coincidence probability contours as a function of flux density and $R$ in the lower panel of Figure~\ref{fig:nc}. We denote the contours corresponding to $P_{\rm cc}=0.01$ and $0.1$. We also show the flux density of the persistent radio source associated with FRB\,121102 over the redshift range $z \approx 0.1 - 1$, as well as the nominal sensitivities for a number of existing and future radio observatories. We note that the FRB 121102 persistent source falls well below $P_{\rm cc} = 0.01$. While most existing facilities are sensitive to FRB\,121102-like persistent sources across a range of redshifts ($z \approx 0.1-1$; see Table 1), detecting such a source at $z\sim 1$ (with an expected flux density of a few $\mu$Jy) is challenging even with the VLA. On the other hand, the advent of the ngVLA and SKA will push the achievable sensitivity to sub-$\mu$Jy levels, and hence the detection of such persistent sources to $z\gtrsim 1$.

At $\lesssim 1$ mJy (i.e., FRB\,121102-like persistent sources at $z\gtrsim 0.1$), confident associations $(P_{\rm cc} \lesssim 0.01)$ require localizations of $R \lesssim 20''$. At higher redshifts ($z \sim 1$), localizations of $R \lesssim 1''$ are required. While this level of localization is not feasible for most FRB search telescopes, it can be achieved using the VLA in an extended configuration, or with very long baseline interferometry (VLBI), as in the case of FRB\,121102, which was localized to $\lesssim 12$ mas \citep{Marcote2017}. However, a number of facilities, including the VLA, MeerKAT, ASKAP, and DSA-10, will be able to provide the $\lesssim 20''$ localizations required for robust associations with sub-mJy sources at moderate redshifts ($z \sim 0.5$). Conversely, at $R \gtrsim 1'$, confident associations with sub-mJy sources are not feasible. This corresponds to the localization regime for CHIME, UTMOST-2D and Apertif (as well as single-dish telescopes). These facilities can provide robust associations only if the counterparts generally have flux densities of $\gtrsim 1$ mJy, but this would require some of the persistent radio counterparts to be much more luminous than the source associated with FRB\,121102. Finally, the poor localizations from single dish telescopes are not sufficient for robust associations with all but the brightest ($\sim {\rm Jy}$) sources. Although these telescopes may reveal additional bursts from FRBs in the case of repetitions, they will not be able to directly provide localizations that will lead to associations with persistent radio sources at any reasonable confidence level. 

In Table 1, we list a number of radio facilities designed to detect FRBs. We sort these by anticipated or known localization capability. We also list the flux density of a radio source in the respective localization region that would have $P_{\rm cc}=0.01$ and 0.1, as well as the maximum redshift $z_{\rm max}$ out to which an association with an FRB\,121102-like persistent source can be made. These values are extracted directly from the probability contours in Figure~\ref{fig:nc}. We find that while the VLA, ASKAP, DSA-10, and MeerKAT are capable of probing these sources out to $z\sim 1$ (and $z >1$ for VLBI), the large flux densities required for $P_{\rm cc} \approx 0.01$ for facilities with localizations of $R\gtrsim 1'$ preclude associations with these sources at $z\gtrsim 0.1$. Given the expectation from radio source number counts and the sensitivity of current instruments, localizations below $3''$ will not improve the association confidence markedly, as existing facilities cannot achieve the $\mu$Jy levels required for $P_{\rm cc} \lesssim 0.1$. 

We note that for the same localization requirement of $R\lesssim 20''$, optically-based host galaxy associations are impractical, requiring host luminosities several times brighter than $L^*$ at redshifts below $z \lesssim 0.1$ (Paper I). Assuming that FRB\,121102 is representative of the FRB population as a whole and that we can expect continuum radio sources coincident with FRBs, the lower areal number density of radio sources on the sky enables more robust associations at a given localization precision; we demonstrate this further using optical and radio observations of the localization region of FRB\,170107 in \S\ref{sec:example}.

\begin{figure*}
\includegraphics[width=\textwidth]{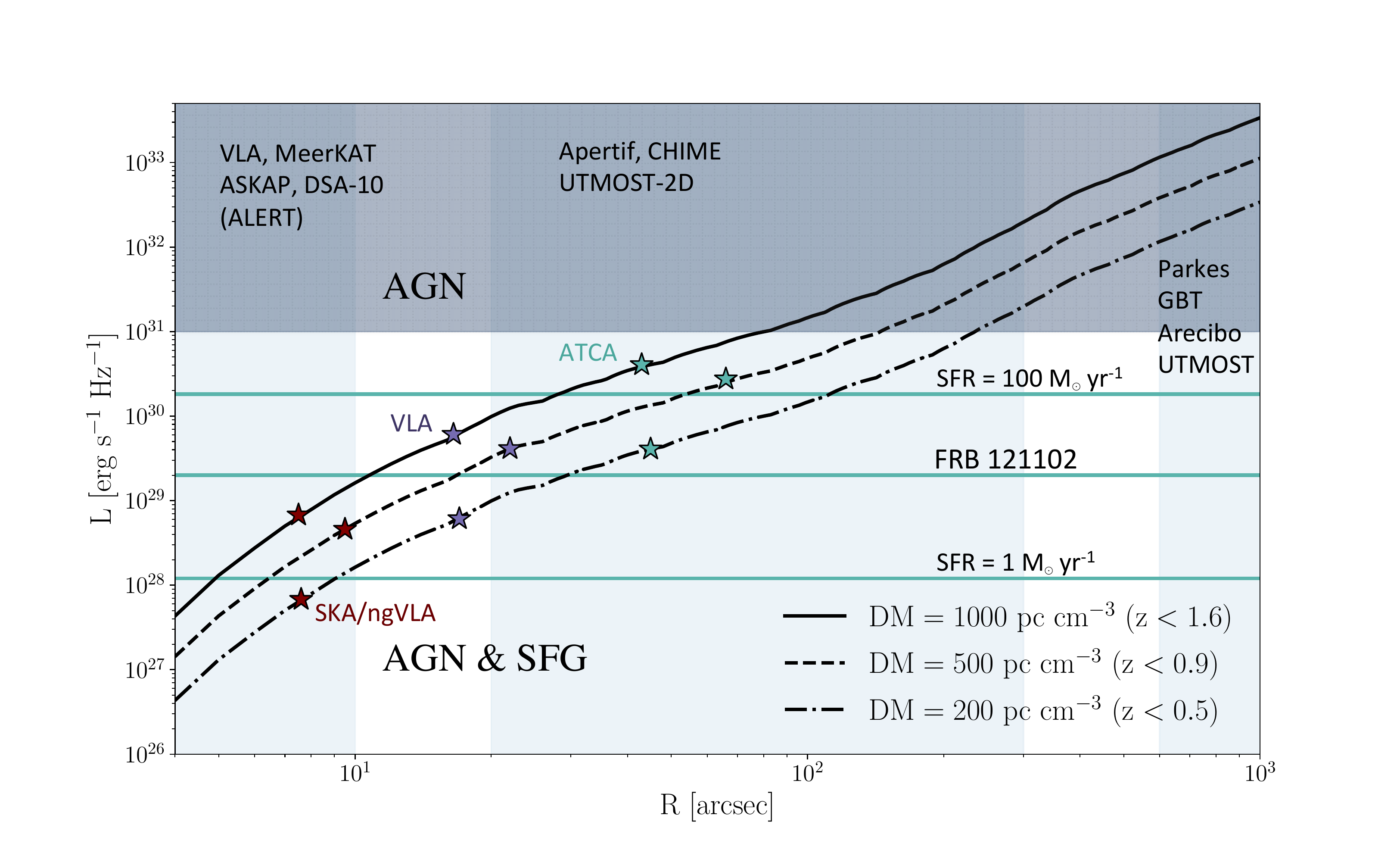}
\caption{The upper limit on the luminosity of an associated radio source as a function of localization radius for a range of DM values.  The limit is calculated using the number counts in Figure~\ref{fig:nc} set to an expectation value of one source and the upper limit on the redshift for each DM value. We plot as horizontal lines the luminosity of the quiescent radio source associated with FRB\,121102 and the luminosities for star forming galaxies with star formation rates of 1 and 100 $\rm M_{\odot} \ yr^{-1}$. We also indicate the dividing line above which radio sources are almost all AGN ($L \approx 10^{31} \rm \ erg \ s^{-1} \ Hz^{-1}$). Vertical bars indicate the localization regimes of various radio telescopes that are designed to or capable of detecting FRBs (Table 1). Also shown are the achievable luminosity limits (for each DM, or equivalently $z_{\rm max}$) for follow-up observations with ATCA, VLA, and the SKA/ngVLA.}
\label{fig:ul}
\end{figure*}

\section{Placing an Upper Limit on the Radio Luminosity}\label{sec:ullim}
We use the results of \S\ref{sec:nc} to investigate upper limits on the radio luminosity of an associated persistent source in the absence of a robust association. For a given localization radius, we determine the typical brightest source expected within the region (using the top-right panel of Figure~\ref{fig:nc}). For a given intergalactic DM value (excluding the host and Milky Way contribution), we then estimate $z_{\rm max}$, using the DM-redshift relation of \citet{Deng2014} (see also \citealt{Ioka2003}) and incorporating the uncertainty due to IGM inhomogeneities as parameterized by \citet{McQuinn2014}. We follow the procedure described in detail in Paper I. In Figure~\ref{fig:ul}, we plot the maximum luminosity of a radio source as a function of $R$ for a range of DM values. As in Figure~\ref{fig:nc}, we overlay the localization capabilities of the various FRB search facilities for reference. We also denote lines corresponding to the luminosity of star-forming galaxies with star formation rates of 1 and 100 $\rm M_{\odot} \ yr^{-1}$, and we indicate the dividing line above which radio sources are almost exclusively AGN ($L \approx 10^{31} \rm \ erg \ s^{-1} \ Hz^{-1}$). Finally, we mark the achievable luminosity limits as a function of redshift for several facilities (ATCA, VLA, and the SKA/ngVLA).

\begin{deluxetable}{lcccc}
\tablecolumns{5}
\tablewidth{0pt}  
\caption{Radio facilities and their localization capabilities.}
\tablehead{
\colhead{Telescope} & 
\colhead{$R_{\rm FRB}$} & 
\colhead{$S(P_{\rm cc} = 0.01)$} & 
\colhead{$S(P_{\rm cc} = 0.1)$} & 
\colhead{$z_{\rm max}$} \\ 
\colhead{} & 
\colhead{[arcsec]} & 
\colhead{[mJy]} & 
\colhead{[mJy]} &
\colhead{($P_{\rm cc} = 0.01$)}
}  
\startdata
VLBA / EVN & $0.001\text{--}0.1$ & $<10^{-4\dagger}$ & $<10^{-4\dagger}$ & $>1$\\
VLA & $0.1\text{--}3$ & 0.01 & $<10^{-4\dagger}$ & 0.6 \\
ASKAP & $0.8\text{--}1.5$ & 0.03 & 0.002 & 0.4\\
DSA-10 & $1\text{--}2$ & 0.04 & 0.005 & 0.4 \\
MeerKAT & $2\text{--}10$ & 0.2 & 0.03 & 0.2 \\
UTMOST-2D & $2\text{--}30$ & 2 & 0.2 & 0.1\\
Apertif$^{\ddagger}$ & $5\text{--}60$ & 20 & 0.8 & $z < 0.1$\\
CHIME & $20\text{--}600$ & 200 & 30 & $z < 0.1$\\
UTMOST & $100\text{--}300$ & 400 & 60 & $z < 0.1$\\
Arecibo & $200\text{--}230$ & 500 & 100 & $z < 0.1$ \\
Parkes & $500\text{--}800$ & 1000 & 400 & $z < 0.1$\\
GBT & $500\text{--}800$ & 1000 & 400 & $z < 0.1$\\
\enddata
\tablecomments{Radio facilities capable of detecting FRBs, ordered by approximate localization capability. $\dagger$ Limits are used to denote flux values which extend below the radio source number counts. $\ddagger$ In conjunction with LOFAR, the Apertif LOFAR Exploration of the Radio Transient Sky (ALERT) survey can provide more accurate (arcsecond) localizations (\citealt{vanLeeuwen2014}; see also http://alert.eu/).}
\end{deluxetable}

The results suggest that for DM $\approx 1000 \rm \ pc \ cm^{-3}$, an upper limit comparable to the luminosity of the FRB\,121102 persistent radio source ($L \approx 2 \times 10^{29} \rm \ erg \ s^{-1} \ Hz^{-1}$; \citealt{Marcote2017}) can be placed if $R \lesssim 10''$. These limits would also rule out star formation at the level of $\approx 100 \ \rm M_{\odot} \ yr^{-1}$. A similar localization in the optical would only constrain the host galaxy luminosity to $\approx L^*$ (see Paper I, Figure 3). However, these limits are below the nominal sensitivities for existing radio telescopes. For example, although the VLA can provide the required localization precision, the $3\sigma$ limiting flux density at 6 GHz corresponds to a luminosity upper limit of $L \approx 6 \times 10^{29} \rm \ erg \ s^{-1} \ Hz^{-1}$. For lower DMs, similar constraints on the luminosity of a quiescent radio source can be placed for larger localizations, i.e., $R \lesssim 30''$ for DM $\approx 200 \rm \ pc \ cm^{-3}$. 

Although the upcoming SKA and ngVLA will provide the sensitivities required for meaningful upper limits, similar limits can be placed with existing facilities only for lower DM values (DM $\lesssim 500 \ \rm pc \ cm^{-3}$).

\section{Rejecting Spurious Radio Associations}\label{sec:cands}
While the probability of chance coincidence analysis in \S\ref{sec:nc} provides a direct measure of the confidence level of associating an FRB with a persistent radio source, we are also interested in exploring additional ways of rejecting spurious associations with unrelated radio emission due to AGN or star forming galaxies (i.e., the sources that make up the extragalactic radio source counts).  First, we discuss a number of methods that can be used to identify radio emission due to star formation and separate these sources from FRB host candidates. Next, we discuss the distinction between AGN and putative FRB hosts, which is generally more complicated due to the wide range of AGN radio properties. 

\subsection{Rejecting Star Forming Galaxies}

\begin{figure}
\includegraphics[width=\columnwidth]{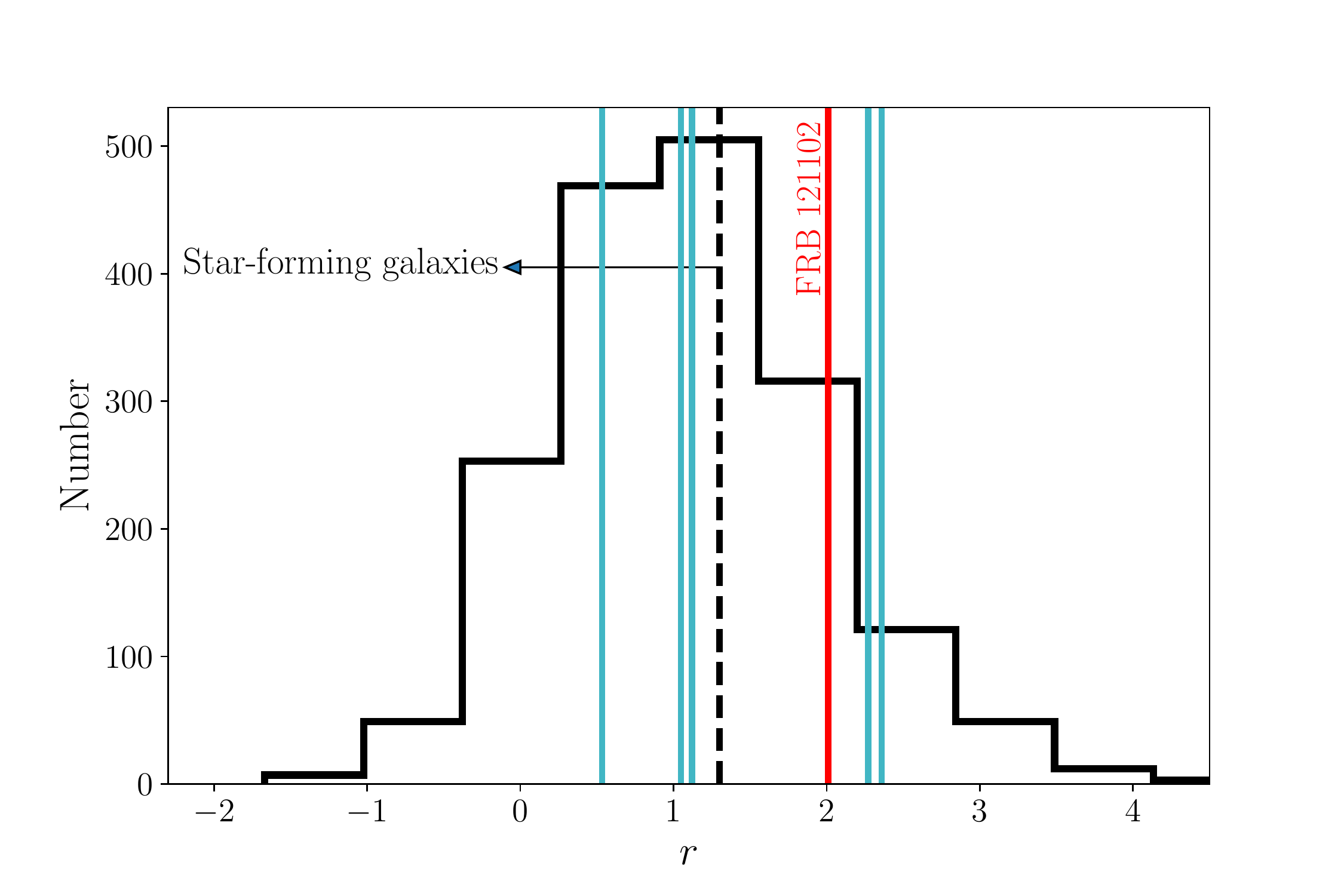}
\caption{Distribution of the radio-to-optical flux ratio [$r = \rm{log}$$(S_{\rm 1.4 GHz}/S_V)$] for 1784 radio sources from the literature (\citealt{Machalski1999}; \citealt{Afonso2005}; \citealt{Padovani2009}). The dashed line indicates the nominal maximum value for star forming galaxies. Blue lines correspond to the five radio sources that we identified in the localization region of FRB\,170107 (see Table 2). Also shown in red is the ratio for the persistent source associated with FRB\,121102.}
\label{fig:fluxratio}
\end{figure}

Motivated by the properties of the FRB\,121102 persistent radio counterpart, the ratio of radio-to-optical flux can be used to rule out radio emission due to star formation. This ratio, commonly defined as $r\equiv \rm{log}$$(S_{\rm 1.4 GHz}/S_V)$, where $S_V$ is the $V$-band flux density, has previously been used as a discriminant between radio-loud AGN and star forming galaxies, where starburst galaxies have $r<1.4$ (\citealt{Machalski1999}; \citealt{Afonso2005}; \citealt{Barger2007}; \citealt{Seymour2008};\citealt{Vega2008}; \citealt{Padovani2009}). In Figure~\ref{fig:fluxratio}, we plot the distribution of radio-to-optical flux ratios for radio sources from the literature (\citealt{Machalski1999}; \citealt{Afonso2005}; \citealt{Padovani2009}) and for the FRB\,121102 persistent source \citep{Chatterjee2017}. In the case of FRB\,121102, the presence of bright radio emission, coupled with a faint optical host, leads to $r\approx 2.0$ and implies that the radio emission does not arise from star formation. Assuming that this is generally the case for FRB counterparts, we can set a threshold of  $r\gtrsim 1.4$ which effectively reduces the source counts by a factor of about two. 

The angular extent of the radio emission can also be used to distinguish star forming galaxies from compact FRB radio counterparts \citep{Ofek2017}. The radio emission from the latter will have a scale of $\lesssim 1$ pc, as in the case of FRB\,121102 ($\lesssim 0.7$ pc; \citealt{Marcote2017}), appearing as unresolved sources. Star forming regions at this scale would point to extremely high star formation rates per unit area ($\sim 10^{5-7} \ \rm M_{\odot} \ yr^{-1} \ kpc^{-2}$; \citealt{Park2016}). This level of star formation activity is expected in only the most extreme star forming regions \citep{Varenius2014}. This argument is borne out by VLBI observations, which show that about $80\%$ of radio sources brighter than 1 mJy are resolved, thereby implicating extended regions of star formation or FRII AGN \citep{Deller2013}. Thus, VLBI follow-up of any candidate FRB counterparts can be used to reject extended radio sources.

\subsection{Rejecting Active Galactic Nuclei}
The wide range of radio emission properties in AGN makes their rejection more challenging. For example, the radio-to-optical flux ratios for AGN exhibit a range of values. \citet{Keane2016} claimed the first precise localization of an FRB based on the detection of a contemporaneous radio counterpart within the Parkes Telescope localization region. The counterpart was subsequently shown to be a variable AGN undergoing strong refractive scintillation \citep{Williams2016}. A search for counterparts in the FRB\,131104 localization region similarly revealed a coincident variable AGN \citep{Shannon2017}. The approach we advocate here is to use various multi-wavelength data to argue whether a source is consistent with an AGN, and hence not securely associated with an FRB. 

Optical emission lines can be used to identify AGN using the Baldwin, Phillips $\&$ Terlevich (BPT) diagram \citep{Baldwin1981}. However, if we impose a high radio-to-optical flux ratio, then obtaining optical spectra of the associated host galaxies at $z \gtrsim 0.5$ will be beyond the reach of most ground-based facilities.

Sources that are precisely coincident with the nuclei of their host galaxies are more likely to have an AGN origin. However, this is less clear in the case of dwarf galaxies in which the radio source may be offset from the optical center of the galaxy \citep{Reines2011} and the optical center may not be well defined. The fraction of dwarf galaxies which host AGN is thought to be quite small, however, with an occupation fraction of 0.5$-$3\% (\citealt{Reines2013}; \citealt{Pardo2016}). 

\begin{deluxetable}{llccc}
\tablecolumns{4}
\tablewidth{0pt}  
\caption{Radio sources in the FRB\,170107 localization region.}
\tablehead{
\colhead{RA} & 
\colhead{DEC} & 
\colhead{Flux Density} &
\colhead{$m_i$} &
\colhead{$r$} \\ 
\colhead{} & 
\colhead{} & 
\colhead{[$\microJy$]}
}  
\startdata
11:23:14.739 & $-$04:58:20.47 & 86$\pm$15 & 20.4 & 0.5\\
11:23:24.431 & $-$04:59:33.21 & 112$\pm$15& 21.4 & 1.0\\
11:23:20.400 & $-$05:00:38.59 & 76$\pm$14 & 22.0 & 1.1\\
11:23:11.227 & $-$04:57:14.38 & 129$\pm$23 & $>24.3$ & $>2.3$\\
11:23:26.385 & $-$04:57:41.65 & 158$\pm$16 & $>24.3$ & $>2.4$\\
\enddata
\tablecomments{Optical non-detections are listed as the $3\sigma$ limiting magnitude.}
\end{deluxetable}

\begin{figure*}
\includegraphics[width=\textwidth]
{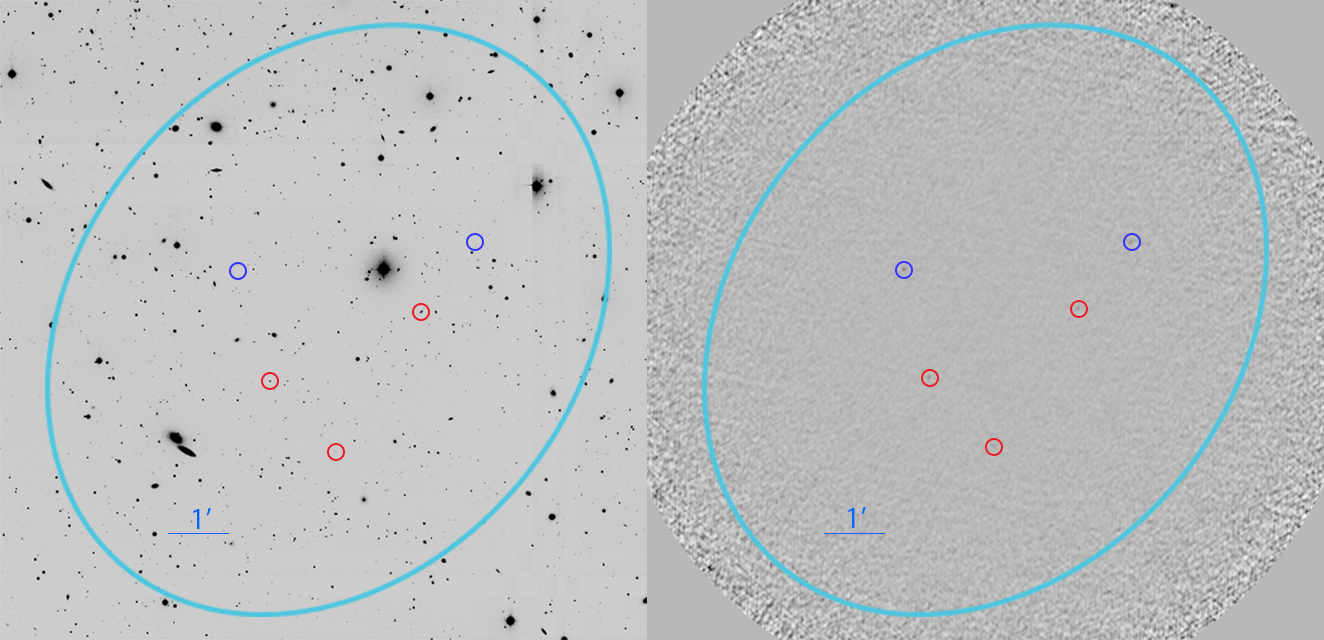}
\caption{Optical (left) and radio (right) images of the $90\%$ ASKAP localization region for FRB\,170107 (\textit{cyan}) with a pointing center given by $\text{RA} = 11{:}23{:}21$, $\text{Decl.} = -04{:}58{:}38$. A total of five sources are identified in the radio image. Blue circles correspond to candidate counterparts based on high radio-to-optical flux ratios; red circles correspond to sources with ratios typical of star forming galaxies.}
\label{fig:image}
\end{figure*}

Finally, X-ray observations can be used to discern AGN activity, with X-ray luminosities of $L_X \gtrsim 10^{42} \rm \ erg \ s^{-1}$ indicative of AGN, although low-luminosity AGN may exhibit lower values. In the case of FRB\,121102, X-ray emission is not detected, corresponding to a $3\sigma$ upper limit of $L_X \lesssim 3 \times 10^{41} \rm \ erg \ s^{-1}$ at 0.5$-$10 keV \citep{Chatterjee2017}. The possibility of a low-luminosity, radio-loud AGN cannot be precluded from this result, however. In general, the absence of detectable X-ray emission may not be constraining, while a direct detection can be used to argue for an AGN origin. 

\section{A Practical Example: FRB\,170107}\label{sec:example}

FRB\,170107 \citep{Bannister2017} was the first FRB detected by the Australian Square Kilometer Array Pathfinder (ASKAP; \citealt{Johnston2008}; \citealt{Schinckel2012}) using the phased array feed system. This led to the detection of the pulse in three separate beams, allowing for a localization region smaller than the size of an individual beam: the 90\% confidence ellipse has semi-major and semi-minor axes of $5.3' \times 4.2'$ with a position angle of $52^\circ$. Motivated by the smaller size of the localization region compared to previous FRBs and the analysis presented above, we obtained radio and optical follow-up observations shortly after publication of the event to search for a host galaxy via a radio source with a large radio-to-optical flux ratio.

\subsection{Radio Observations}

We obtained radio observations with the Karl G. Jansky Very Large Array (VLA\footnote{The VLA is operated by the National Radio Astronomy Observatory, a facility of the National Science Foundation operated under cooperative
agreement by Associated Universities, Inc. The observations presented here were obtained as part of program VLA/17A-453.}) on 2017~Jun~2, 146 days after the FRB. The observation lasted 30 minutes and the pointing center was $\text{RA} = 11{:}23{:}10$, $\text{Decl.} = -05{:}01{:}00$, matching the FRB position centroid reported by \citet{Bannister2017}. The correlator was configured in the standard ``C band'' wideband continuum mode using the 3-bit samplers, covering the full bandwidth of 4--8~GHz. Standard calibration techniques were used. The bandpass and flux density calibrator was 3C\,286 and the complex gain calibrator was the blazar PKS~J$1131{-}0500$.

After this observation was conducted, we were provided with an updated localization probability map with a new centroid that differed from the published position (R. Shannon and K. Bannister, 2017, priv. comm.). On 2017~Jun~22 we conducted a second observation using the improved pointing center of $\text{RA} = 11{:}23{:}21$, $\text{Decl.} = -04{:}58{:}38$. The two positions are separated by almost exactly the half-width at half power (HWHP) of the VLA at 6~GHz. The other characteristics of this observation were identical to those of the first.

We calibrated and imaged both observations within the CASA software environment \citep{McMullin2007} using standard techniques. We imaged the fields separately, creating images of $1024{\times}1024$ pixels at a scale of 1$''$ per pixel using multi-frequency synthesis \citep[MFS;][]{Sault1994} and $w$-projection with 128~planes \citep{Cornwell2008}. The images of the first and second pointing centers achieved rms values of 8 and 10~\microJy\ at the pointing center, respectively.

We visually identified radio sources and measured their flux densities following a primary beam correction. Due to the decrease in sensitivity away from the pointing center, we impose a 75 $\mu$Jy threshold and find five radio sources above this limit. The radio image from our second observation (2017 Jun 22) is shown in Figure~\ref{fig:image} and the identified sources are listed in Table 2.

\subsection{Optical Observations}
We obtained $i$-band observations covering the FRB localization region with IMACS on the 6.5 m Magellan Baade Telescope. We processed the images using standard procedures in {\tt IRAF} and calibrated the resulting magnitude measurements using field stars in common with the Pan-STARRS1 $3\pi$ survey.  We identify optical counterparts to three of the five radio sources (see Figure~\ref{fig:image}). We list the apparent $i$-band magnitudes (corrected for Galactic extinction\footnote{http://irsa.ipac.caltech.edu/applications/DUST/}; \citealt{Schlafly2011}) for each source in Table 2. The two optical non-detections correspond to a 3$\sigma$ limiting magnitude of 24.3 mag.

\subsection{Candidate Sources}

Based on the flux densities of the detected sources ($76-158 \ \mu$Jy), we estimate the expected number of sources above 75 $\mu$Jy from the radio source number counts. To compare our sources at 6 GHz to the expectation from the 1.4 GHz source counts, we assume a spectral index of $\alpha = -0.7$ \citep{Condon2012}, corresponding to an effective limiting flux density of 210 $\mu$Jy at 1.4 GHz. From our analysis in \S\ref{sec:nc}, we therefore expect to find $\sim 7$ radio sources in the localization region, consistent with our results. The flux densities of the detected sources preclude a robust association, however. Namely, even for the brightest detected source, $P_{\rm cc} \approx 0.99$; $P_{\rm cc} = 0.01$ requires a $\approx 0.1$ Jy source. 

Motivated by the persistent radio source associated with FRB\,121102, we determine the ratio of radio-to-optical flux, $r=\rm log(S_{\rm 6GHz}/S_i)$, for each source. Of the five identified radio sources, two lack optical counterparts, corresponding to large radio-to-optical flux ratios ($r\gtrsim 2.3$ and $r\gtrsim 2.4$), comparable to the value observed for the FRB\,121102 counterpart ($r\approx 2.0$) and well in excess of expected values for star forming galaxies (see Figure~\ref{fig:fluxratio}). Conversely, the three sources with optical counterparts have $r\lesssim 1.1$, within the regime of star forming galaxies. We therefore consider the two radio sources with optical non-detections as potential hosts of FRB\,170107, given that we can rule out star formation as the source of the radio emission. We refer to the 158 and 129 $\mu$Jy sources as G1 and G2, respectively.

Assuming a nominal host DM contribution of $\sim 100 \ \rm pc \ cm^{-3}$ (as in \citealt{Bannister2017}), we estimate an upper bound on the redshift using the same approach as in \S\ref{sec:ullim}. We find that the DM-inferred redshift can be characterized as a normal distribution with $\langle z\rangle\approx 0.54$ and a 68\% confidence range of $0.18-0.90$. The range of possible luminosities is therefore $1 \times 10^{29} - 4 \times 10^{30} \ \rm erg \ s^{-1} \ Hz^{-1}$, comparable to the luminosity of the FRB\,121102 persistent source ($L \approx 2 \times 10^{29} \rm \ erg \ s^{-1} \ Hz^{-1}$; \citealt{Marcote2017}). The mean luminosities (assuming $z \approx 0.54$) are $1.3 \times 10^{30} \rm \ erg \ s^{-1} \ Hz^{-1}$ and $1.0 \times 10^{30} \rm \ erg \ s^{-1} \ Hz^{-1}$ for G1 and G2, respectively.

The $3\sigma$ limiting magnitudes in the optical ($m_i > 24.3$) correspond to $\approx 0.1$ $L^*$ at $z\approx 0.54$ (or $\approx 0.01$ $L^*$ at $z\approx 0.18$ and $\approx 0.3$ $L^*$ at $z\approx 0.90$). These are broadly consistent with the host galaxy of FRB\,121102 which has a luminosity of  $\approx 0.01$ $L^*$ \citep{Tendulkar2017}.

As both the radio and optical properties of our candidate sources are consistent with the properties of the FRB\,121102 host galaxy and radio counterpart, we argue that these sources are viable counterparts to FRB\,170107. The large localization precludes a robust association with either source, however, given that the expectation value for such a source is $> 1$ (\S\ref{sec:nc}). Targeted follow-up observations with radio interferometers may reveal additional bursts from these sources, leading to an unambiguous host identification.

\section{Conclusions}\label{sec:conc}

Motivated by the persistent radio source associated with the repeating FRB\,121102, we explored the likelihood of identifying persistent radio sources associated with FRBs as a function of their flux density and the FRB localization radius, assuming that all FRBs are associated with such sources. We also quantified the limits that can be placed on a radio counterpart in the absence of a robust association. Finally, we applied our analysis to the case of FRB\,170107. Our main results can be summarized as follows:

\begin{itemize}
\item{Localizations of $R \lesssim 20''$ are required for robust associations ($P_{\rm cc}=0.01$) with sub-mJy radio sources at $z \approx 0.1$. At higher redshifts ($z \approx 1$), smaller localizations of $R \approx 1''$ are needed. A number of radio facilities, including the VLBA, EVN, VLA, MeerKAT, ASKAP, and DSA-10 will be capable of providing such FRB localizations. For large localizations ($\approx 300''$), only bright sources ($\approx 10 - 100$ Jy) lead to $P_{ \rm cc} \lesssim 0.01$. However, such sources are not obviously expected given the luminosity of the FRB\,121102 counterpart and the expected extragalactic redshifts. Nevertheless, several radio facilities, including Apertif, CHIME, and UTMOST-2D will be able to achieve this level of accuracy.}

\item{Radio sources with luminosities similar to the FRB\,121102 persistent source beyond $z\sim 0.5$ are not easily detectable with existing facilities, since they will have flux densities of only a few $\mu$Jy. However, the upcoming SKA and ngVLA will be sensitive to such sources beyond $z\sim 1$.}

\item{In the absence of a robust association, an upper limit can be placed on the radio luminosity of an associated radio source. These limits are particularly interesting for localizations of $R \lesssim 10''$ (for DM $\approx 1000 \rm \ pc \ cm^{-3}$) or $R\lesssim 30''$ (for DM $\approx 200$) where the luminosity is comparable to that of the FRB\,121102 persistent source. Conversely, in the optical, a $10''$ localization does not place a particularly meaningful constraint on the host galaxy luminosity (upper limit of $L^*$ for $\rm DM \approx 1000 \rm \ pc \ cm^{-3}$; Paper I). These limits are below the imaging capabilities of existing facilities, however, except for low DM values (DM $\lesssim 500 \ \rm pc \ cm^{-3}$).}

\item{A number of multi-wavelength techniques can be used to reject spurious associations with unrelated radio sources, namely AGN and star forming galaxies. The radio-to-optical flux ratio can be used to reject star forming galaxies as FRB host candidates, imposing a threshold of $r \gtrsim 1.4$ for FRB counterparts. Although the rejection of AGN is generally more challenging, optical emission lines and X-ray observations can be leveraged to determine whether a source is consistent with an AGN. }

\item{Using a combination of radio and follow-up observations, we identified five radio sources above 75 $\mu$Jy within the FRB\,170107 localization region. Two of these sources share properties consistent with the FRB\,121102 persistent source, namely a high radio-to-optical flux ratio, a comparable radio luminosity, and a low luminosity optical counterpart. We find that the range of allowed radio luminosities ($1 \times 10^{29} - 4 \times 10^{30} \rm \ erg \ s^{-1} \ Hz^{-1}$) is comparable to that of the FRB\,121102 persistent source ($2 \times 10^{29} \rm \ erg \ s^{-1} \ Hz^{-1}$). Similarly, the optical limits ($m_i > 24.3$ mag) are consistent with the dwarf host galaxy of FRB\,121102 ($m_r \approx 23.3$ mag). }

\end{itemize}

Combined radio and optical follow-up observations of FRB localization regions can provide a powerful tool for identifying host associations. If FRB\,121102 is representative of the FRB population, candidate sources can be identified on the basis of their radio-to-optical flux ratio, and targeted follow-up observations can be used to search for repetitions directly from these sources. Although continued monitoring with ASKAP or single dish telescopes may reveal additional bursts from FRBs, these will not provide unambiguous host identifications. Detections of additional bursts with targeted phased-array observations with the VLA would not only demonstrate repetitions, confirming the existence of a class of repeating FRBs, but would also immediately lead to a host identification. This technique may therefore provide a novel framework for identifying counterparts and localizing future FRBs. However, in the absence of repetitions or very precise localizations, definitive associations remain challenging, particularly due to the broad region of parameter space occupied by AGN.

\acknowledgments \textit{Acknowledgments.} The Berger Time-Domain Group at Harvard is supported in part by the NSF through grant AST-1714498, and by NASA through grants NNX15AE50G and NNX16AC22G. This research has made use of the FRB Catalogue ({\tt \footnotesize{http://www.astronomy.swin.edu.au/pulsar/frbcat/}}) and NASA's Astrophysics Data System.

\software{CASA \citep{McMullin2007}, pwkit \citep{Williams2017}}

\appendix
\section{Accounting for the Primary Beam Response}\label{sec:pb}

In \S\ref{sec:nc}, the chance coincidence probability is calculated assuming uniform sensitivity over a region with a 1$\sigma$ localization radius given by $R_{\rm FRB}$. However, in typical radio observations, sensitivity is not spatially uniform. Consequently, the number density of detectable point sources decreases. In the common case of a Gaussian primary beam, the limiting flux density is given by $S_{\rm lim} = S \cdot e^{-r^2/2 \sigma_{\rm PB}^2}$, where $r$ is the distance from the pointing center, $\sigma_{\rm PB} \approx \theta_{\rm PB}/2.4$, and $\theta_{\rm PB}$ refers to the FWHP (full-width at half power) of the primary beam. For a given localization radius $R$ and flux density $S$, the resulting chance coincidence probability $P_{\rm cc}$ therefore increases. In this scenario, the chance coincidence probability is given by:
\begin{equation} 
P_{\rm cc} = 1 - e^{-\int_0^{R} 2 \pi r \ n(S \cdot e^{-r^2/2 \sigma_{\rm PB}^2}) \ dr} .
\end{equation}

\begin{figure}[!ht]
\centering
\includegraphics[width=10cm]{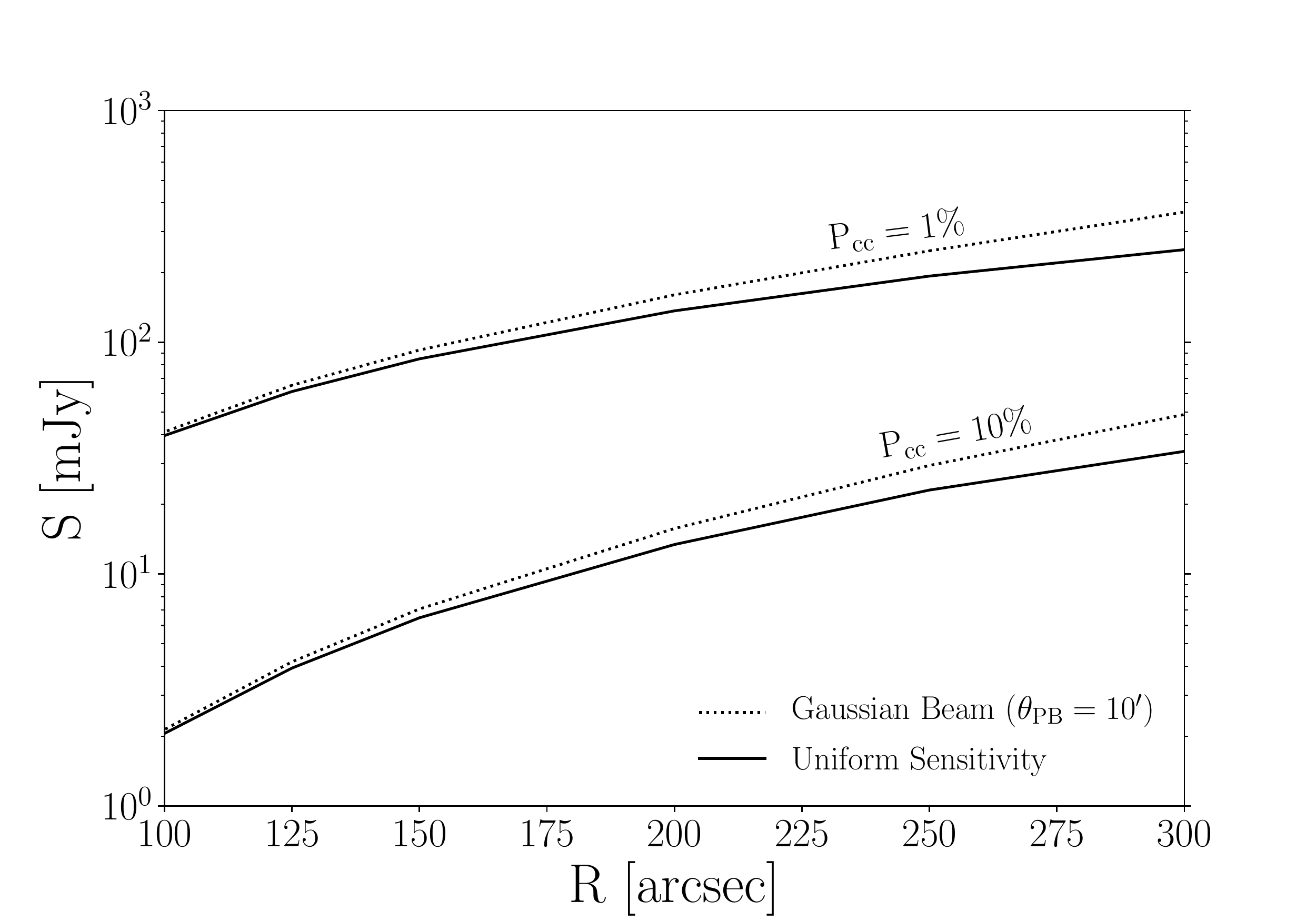}
\caption{Comparison of $1\%$ and $10\%$ probability contours assuming a uniform-sensitivity beam and a Gaussian primary beam with $\theta_{\rm {PB}} = 10'$.}
\label{fig:beam}
\end{figure}

We illustrate the effects of accounting for the primary beam response in Figure~\ref{fig:beam}. Namely, we recalculate the chance coincidence probability using Equation A1, where we use, as an example, $\theta_{\rm PB} = 10$ arcminutes (representative of the VLA at 4.5 GHz). We find that in this regime, an accurate characterization of the beam produces results that are comparable to the uniform sensitivity case across the extent of the primary beam ($R = 300''$). For example, at $R=300''$, using Equation 1, we find that a $1\%$ chance coincidence probability occurs for a source flux density of $\approx 250$ mJy. Conversely, in the case of a Gaussian beam, the same flux density corresponds to a $\approx 1.2\%$ chance coincidence probability, where $P_{\rm cc} = 1\%$ requires $S \approx 300$ mJy. We therefore conclude that our simplified beam-model analysis in \S\ref{sec:nc} is sufficient in terms of characterizing $P_{\rm cc}$ and identifying the required localization precision for confident associations.

\bibliography{ref}
\bibliographystyle{apj}

\end{document}